# Combining Absolute and Relative Pointing for Fast and Accurate Distant Interaction

Yunfeng Zhang
IBM T. J. Watson Research Center
zhangyun@us.ibm.com

**ABSTRACT**
Traditional relative pointing devices such as mice and trackpads are unsuitable for pointing at distant displays, because they encumber the users by requiring either a flat surface to operate on or being held by two hands. Past research has examined many new pointing methods, but few could surpass the speed and accuracy of mice and trackpads. This paper introduces a new pointing system that is developed based on HTC Vive, a relatively low-cost virtual reality system, and proposes two methods of combining absolute and relative pointing. The proposed methods were compared against single-mode pointing methods (i.e., pure absolute pointing and pure relative pointing) in a Fitts' law study. The results show that with only a short period of practice, one hybrid pointing technique enabled faster and more accurate pointing than both single-mode pointing techniques, which included a trackpad.

**ACM Classification Keywords**
H.5.2. [Information Interfaces and Presentation]: User Interfaces — Input devices and strategies.

**INTRODUCTION**
Today, many collaborative environments such as classrooms and meeting rooms have very large displays for presenting visual materials. These displays act as shared workspaces so that everyone in the room can easily refer to contents on the display or see what others are talking about. However, interacting with contents on these large displays is not easy. Traditional pointing devices such as mice and trackpads require a flat surface to operate on, which inhibits people from moving freely to engage with others. In addition, when there are displays at different directions, it can be difficult to find the cursor and it can also become unclear as to how to move the cursor from one display to another. Large displays in such environments are also not conducive to touch interactions because (a) people have to move close to the displays to interact with them, and (b) parts of the displays may be unreachable.

When using large displays, people seem to be more used to absolute pointing devices such as laser pointers than relative pointing devices like mice. Laser pointing is often used in lectures and presentations. It has a couple of advantages over relative pointing: (a) It can be held with one hand and does not require a flat surface to sit on, and (b) people can directly point to somewhere without needing to first find the cursor and then gradually move it to the desired location. Some research [14,15] even made it possible to use laser pointing to directly interact with contents on the display by detecting the laser spot through a computer vision system. A major problem with laser pointing, however, is that the laser spot is virtually undetectable on a LCD screen, which are becoming increasingly popular in today's classrooms and meeting rooms.

Until very recently, inexpensive and accurate absolute pointing devices are difficult to find. One widely known and researched absolute pointing device is the Wii Remote. It is a relatively inexpensive system, and some research (e.g., [9]) explored using it as a distant pointing device. However, the Wii Remote has a very limited range: It only works when the remote is 3 to 10 feet in front of its sensor bar and is somewhat pointed towards the bar [13]. When the display or the room is large, or when there are displays in many different directions, the Wii Remote will not work. Another commercially available pointing system is the Oblong wand. It uses a combination of ultrasound and radio frequency signals to track the wand location and orientation [16]. The tracking is very accurate and the tracking range can be extended by installing more signal emitters on the ceiling. However, its installation is difficult since it requires setting up a grid of signal emitters. In addition, it can be only purchased as a part of a larger system that costs hundreds of thousands of dollars. In research, a more often used tracking system is the VICON motion-capture system, and researchers often create their own pointing devices by attaching IR retroreflective markers to an existing device to track its location and orientation (e.g., [7,12,20]). The VICON system, however, also costs hundreds of thousands of dollars, and is thus not an effective solution for regular consumers or even schools and small companies.

Besides the shortage of inexpensive and accurate devices, there are also other inherent issues to absolute pointing. Prior research has shown that when the target is small, distant absolute pointing has very low accuracy even though the device itself is accurately tracked [5,10,15,20]. This low accuracy is because (a) the human hands has slight natural tremors that are amplified by the distance between the user and the screen, and (b) the pointing device moves slightly when the user clicks a button on the device and this slight movement is again amplified on the remote display. To address this problem, some research proposed methods that combined absolute and relative pointing [5,11,20], but few exceeded the speed of relative pointing [11].

In this paper, I introduce a new absolute pointing device based on a relatively inexpensive virtual reality gaming system, and propose two dual-mode pointing techniques that combine absolute and relative pointing. I evaluated these two pointing techniques against pure absolute and pure relative pointing using a classic Fitts' reciprocal tapping task. The results show that one of the dual-mode pointing techniques can achieve faster selection time than all other techniques, including relative pointing, while still being one of the most accurate methods. This technique also received the highest usability score from the participants. Its inexpensiveness and superb pointing performance makes it an ideal system for interacting with large, distant displays.

**RELATED WORK**

*Head and gaze pointing*. Many studies have investigated using head and gaze pointing for interacting with conventional monitors [22] and very large displays [11,17]. A clear advantage of head and gaze pointing is that they directly reflect people's visual attention (though head direction to a less degree), which is typically where people want to point at. One problem with head and gaze pointing is that they lack a clear selection mechanism, which is also why prior studies often combine them with a manual selection technique. Another problem is that to date, accurate head and eye tracking still relies on the user wearing many sensors such as head-mounted eye trackers and IR retroreflective markers, which is, to some degree, the antithesis of what these techniques are trying to achieve: To free users from encumberment. Thus, these techniques are perhaps still premature for real world applications.

*Freehand pointing*. Freehand pointing is perhaps the holy grail of pointing interaction. Vogel and Balakrishnan [20] pioneered the research in this field. They used passive markers attached to a glove to track hand gestures and pointing directions. They devised different gestures for moving the cursor, performing clicks, and switching between relative and absolute pointing modes. However, just like head and gaze pointing, there is still no reliable, glove free, far range hand tracking technology (though see [18] for recent advancement in vision-based hand tracking research).

*Filtering and smoothing*. Some studies have investigated using digital filters to reduce the effect of hand tremors and device tracking noises. [3,14] applied Kalman filters to remove noises from laser pointers and seemed to have achieved good results. [20] applied a dynamic low-pass filter to hand pointing locations that in effect works similarly to the Kalman filter by adjusting its smoothing factor based on the cursor's moving speed. Though these filters can effectively reduce noise errors and oscillations, they cannot completely remove them. Nor can they reduce the click-induced shakes since those movements are not periodic. When the display is distant, even very small movements would be amplified and hence solutions in addition to filtering are needed.

*Dual-mode pointing*. Several studies have examined the idea of combining some type of relative pointing with pure absolute pointing to increase accuracy. There are roughly two types of dual-mode pointing methods depending on the type of relative pointing integrated. In this paper, I refer to one of them as *hybrid pointing*, which uses finger or hand movements on a plane (such as a trackpad) as the control signal for relative pointing [11,20], and the other as *dual-speed pointing*, which uses the angular change of the absolute pointing direction or its derived signal (such as its intersection point on the display) as the control signal for relative pointing [5,12]. Conceptually, hybrid pointing involves a bigger mode change than dual-speed pointing because with dual-speed pointing, the user would use the same action (shifting the absolute pointing direction) to control the cursor, but with hybrid pointing, a different device and action is usually needed. Nancel et al. [11] compared these two types of dual-mode pointing implemented on very different devices (hybrid pointing was based on head tracking and an iPad, and dual-speed pointing was based on an absolute pointer), and found that they have similar speeds, but hybrid pointing had better accuracy when the target was small. This is because in hybrid pointing, the click action in the relative mode is completely decoupled from cursor position control, whereas dual-speed pointing still suffers from click-induced controller movement that causes the cursor to shift at the moment of clicking. In this research, I implement both dual-mode pointing techniques on the same device, and evaluate them against pure absolute pointing and pure relative pointing.

*Adaptive pointing*. While dual-mode pointing typically requires an explicit action to switch between the pointing modes, adaptive pointing is a technique that implicitly switches the pointing modes based on the moving speed of the pointing device [5,6]. Adaptive pointing is very similar to dual-speed pointing in that it also uses the pointing direction change as the signal for relative pointing. However, the control-display (CD) gain used for relative pointing is gradually adjusted based on the moving speed rather than a fixed constant typically used in dual-speed pointing. It has been shown that adaptive pointing is faster and more accurate than Kalman filter enhanced absolute pointing [6], but is slower than relative, hybrid, and dual-speed pointing [11]. Because of this, I do not evaluate adaptive pointing in this research.

*Control-display transfer function*. One important concept related to relative pointing is the control-display (CD)

transfer function, which transforms the input control signals into cursor displacements on the display. It has been shown that a dynamic transfer function that applies small CD gains for slow input movements and large CD gains for fast input movements performs faster and more accurate than a constant CD gain [1,2]. [11] proposed a sigmoid transfer function and a parameter calibration procedure for large displays. This method is applied to all relative pointing involved in this research.

*Fitts' law and evaluation measures*. It has been shown that both absolute pointing with a fixed screen distance, and relative pointing with constant and dynamic CD transfer functions, follow Fitts' law [2,7]. Because of this, the evaluation measures that apply to Fitts' law can be used to compare the pointing techniques. In this research, I use the widely accepted form of Fitts' equation to model the data:

$$MT = a + b\, log_2\left(\frac{A}{W} + 1\right)$$

where *MT* is the movement time, *a* and *b* are the intercept and slope parameters to be fitted to the data, *A* is the movement amplitude, and *W* is the target width. The log component is also referred to as the index of difficulty (ID). I use *a* and *b* rather than the throughput measure *ID/MT* as the evaluate measure, since it is found that the single throughput measure cannot be generalized to different target widths and distances [21].

## A POINTING DEVICE BASED ON HTC VIVE

To address the need for accurate, long-range, and inexpensive pointing devices, I propose a new pointing system implemented based on HTC Vive, which is a commercially available virtual reality (VR) gaming system. Being a VR system, Vive needs to be able to accurately track where the user is and where he or she is looking. Vive's solution to this problem is similar to Wii Remote's solution, in which infrared light signals are used for devices to locate themselves relative to the light source. The main difference, however, is that the Wii Remote uses static light sources, whereas Vive uses periodic, sweeping lights. The sweeping lights act like a lighthouse, such that based on the timing that the light signals hit the photosensors on the device, the direction of the light source can be determined. By equipping a device with multiple photosensors, based on the known locations of these sensors and the time the signal is received by each sensor, the exact location of the device can be calculated.

This study uses the controllers included in HTC Vive, shown in Figure 1, as the pointing device. The controller is designed to be held in one hand. The cup shaped ring at the tip of the controller contains 24 photosensors that allow it to accurately track its location. The controller has a two-stage trigger button at the bottom, a clickable round trackpad (40 mm in diameter) on the top, a small menu button near the trackpad, and two grip buttons on the sides. The availability of many types of buttons on this controller makes it a potentially very versatile pointing device. The attached trackpad also makes the controller a very useful tool for studying pointing as it enables the possibility to study relative and absolute pointing with one device.

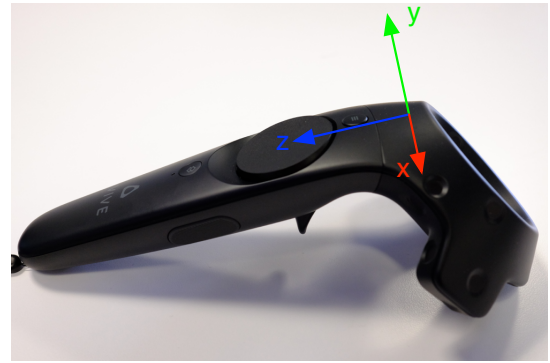

**Figure 1. A Vive controller, superimposed with its native coordinate frame.**

As a pointing system, Vive has several advantages over existing solutions. The Vive tracking system generates data at a relatively high frequency of 90 Hz, can robustly cover a 4 meters by 4 meters area (and can be further expanded at the cost of increasing occlusion likelihood), and has submillimeter tracking noise and under 2 mm tracking error [8]. Compared to the VICON tracking system and Oblong wands, the Vive system is much cheaper ($799) and much easier to deploy since it only involves installing two infrared-laser-emitting stations. Compared to Wii Remote, the Vive system covers much larger area and allows pointing at any directions. Therefore, Vive might be the currently best inexpensive solution as a pointing device for interacting with distant displays.

I developed a C++ program based on the official OpenVR SDK [19] to convert the controllers into pointing devices. The program extracts the controllers' locations, orientations, and button states, encodes the data as a JSON string, and sends the string over network to computers that drive the displays. The following pseudo code describes the key process to extract the controllers' location and orientation.

```
FOR each active controller
  m = getDevicePoseMatrix(controller)
  location = m * Vector(x:0,y:0,z:0,w:1)
  aim = m * Vector(x:0,y:0,z:-1,w:0)
END FOR
```

The controller's location and orientation data are stored in a "device pose" data structure, which is a 4 by 4 affine transformation matrix. The matrix describes the transformation from the device's native coordinate frame, as shown in Figure 1, to the room coordinate frame defined by the laser-emitting stations. Thus, the controller location can be extracted by applying the transformation matrix to the homogeneous coordinate[1] of the origin point. For the pointing direction, note that in the controller's native coordinate system, the pointing direction is (x:0, y:0, z:-1). Therefore, the transformation matrix needs to be applied to the homogeneous coordinate of the negative *z* vector.

On the display side, a nodejs program is written to receive the controller data, and calculate where the controller's pointing direction intersects with the display based on the display's location measured under Vive's room coordinate system. The program then uses the dynamic low-pass filter developed by Vogel et al. [20] to reduce the effect of tracking noises and hand tremors. When the cursor moving speed is low, the filter only allows very low frequency signals to pass (< .25 Hz), and when the cursor speed is high, the filter allows higher frequency signals to pass. As a result, the filter effectively reduces jitters, while also maintains very short lags.

The above description introduces a way to develop an absolute pointing device based on HTC Vive. However, the accuracy of absolute pointing is impaired by hand tremors and click-induced shakes. Even with the dynamic low-pass filter, such effects cannot be completely removed. This paper thus explores two variant pointing techniques that can potentially improve accuracy.

**POINTING TECHNIQUES**

The two variant pointing techniques introduced here are: (a) hybrid pointing, which combines absolute and trackpad-like relative pointing, and (b) dual-speed pointing, which allows the user to slow down the cursor moving speed while still control the cursor in a similar fashion as absolute pointing.

**Hybrid Pointing**

For the hybrid pointing technique, the user switches from absolute to relative pointing by simply touching the controller's trackpad. When in the relative pointing mode, the cursor stops responding to the movements of the controller, and instead responds to the displacements of the user's finger on the trackpad. After the finger is lifted away from the trackpad for 400 ms, the controller switches back to the absolute pointing mode and the cursor quickly and smoothly moves to the absolute pointing location. This delayed mode switch was designed to enable clutching so that the user can keep moving the cursor in the relative pointing mode with multiple swipes on the trackpad.

I used the sigmoid control-display (CD) transfer function proposed by Nancel et al. [11] to determine the cursor movements in the relative pointing mode. This function converts the user's finger moving speed into cursor displacement or CD gain. When the finger moving speed is slow, the CD gain is small, which allows for fine cursor movements. As the finger moving speed increases, the CD gain quickly rises until approaching the predefined maximum CD gain, which allows for fast cursor movement across long distances. The sigmoid transfer function takes six parameters, and I set most of them to the values calibrated by Nancel et al. for their *RelaLarge* trackpad condition, with the exception of the $CD_{max}$ parameter, which was set to 50 for making smaller cursor movements on our wall display.

While previous research has explored similar hybrid pointing ideas, they have some significant differences to the one proposed here. The most similar hybrid pointing techniques are perhaps Vogel et al.'s Hybrid RayToRelative pointing [20] and Nancel et al.'s Head+Coutinuous trackpad pointing [11]. The RayToRelative technique is a freehand pointing technique enabled by passive markers attached to the hand, with the absolute and relative pointing modes associated with different hand gestures. In RayToRelative, relative pointing is done through whole hand movements, which are likely more cumbersome than the finger movements needed here. The Head+Countinous pointing technique uses the head direction, tracked by passive markers attached to the head, as the absolute pointing direction. Relative pointing is completed on a touch device (a tablet or a smartphone) held in the user's hand. Since the gaze direction could not be tracked, the absolute pointing direction is only an approximation of where people are looking, whereas the absolute pointing location of the hybrid pointing technique proposed here is more precise. In addition, both techniques require expensive VICON tracking systems which are mostly inaccessible to regular users. Therefore, the unique form factor and the low cost of the Vive system makes it an ideal testbed for hybrid pointing control.

**Dual-Speed Pointing**

The other technique proposed here is dual-speed pointing, which allows the user to slow down the cursor moving speed by pressing a button. To enable a direct comparison between hybrid and dual-speed pointing, I also designated the trackpad as the mode switch button for dual-speed pointing: when the user touches the trackpad, the cursor enters the slow mode, and when the trackpad is released, the cursor goes back to absolute pointing locations. In the slow mode, the cursor only travels three tenths of the displacement of the absolute pointing location, which facilitates fine cursor movements and drastically reduces the adverse effect of hand tremors and click-induced controller shakes.

The idea of dual-speed pointing may have been first introduced by Kopper et al. [7], who reported that their informal tests showed clear improvements in users' pointing accuracy over pure absolute pointing. A similar idea was implemented as the *LaserGyro* technique in [12], and was found to perform equally well as other top contenders, including pure relative pointing and the Head+Continuous technique [11]. With the LaserGyro technique, a user holds a mouse attached with passive markers for location and orientation tracking. The user switches into the relative mode by pressing and holding the mouse's right button, and makes clicks with the left button. One possible improvement that dual-speed pointing has over LaserGyro is that in dual-speed pointing, the same button is used to perform mode switching (touching the trackpad) and clicks (pressing the trackpad), and because switching to the precise mode is usually what one wants to do before clicking, this design minimizes the cost of mode switch. In this study, I am interested in seeing whether this design would make dual-speed pointing perform significantly better than other techniques.

---

[1] Homogeneous coordinate system adds a fourth coordinate of 1 at the end of 3D coordinates if the coordinates represent a point, and adds a 0 if they represent a vector.

### EXPERIMENT

To determine whether the two dual-mode pointing techniques are better than pure absolute pointing, I conducted an experiment using the Fitts' reciprocal tapping task. My hypothesis is that both hybrid and dual-speed pointing are more accurate than absolute pointing since they allow the user to switch to a more precise pointing mode that reduces and, in the case of hybrid pointing, eliminates the effect of hand tremors and click-induced shakes.

In addition, I was also interested in seeing whether the two dual-mode pointing techniques could outperform pure relative pointing. Thus, I implemented a trackpad-like relative pointing technique and tested it in the experiment. Out of the same consideration as [11] that smartphones and tablets could serve as multifunction controllers for display-rich environments, I implemented this technique on touch devices. More specifically, the relative pointing software was implemented as a webpage that when loaded on a touch device, it would capture the user's finger movements and send the events at 60 Hz to the display-driving machine via a websocket connection. In the experiment, I used an iPad as opposed to a smartphone as the relative pointing device because [11] showed that a larger trackpad area increases performance. On the display side, the same sigmoid transfer function used for hybrid pointing is applied to transform finger movements into cursor displacements. All the parameters of the transfer functions are kept the same except that the $CD_{max}$ was increased from 50 to 200 for faster cursor movements across long distances.

The task involved alternatingly clicking on two vertical bars, and in all techniques that used the Vive controller, clicks were performed by pressing down the trackpad until it clicked, whereas for relative pointing, clicks were done by quick and brief taps on the touch display.

### Participants

18 participants (3 females) were recruited at xxx. All participants had normal or correct to normal vision. One is left handed. 16 of them had prior experience using the Oblong wands, which are absolute pointing devices.

### Apparatus

The experiment stimuli were presented on a wall display consisting of sixteen DynaScan 46'' 1080p thin-bezel LCD monitors arranged into a four by four matrix (see Figure 2). Including the bezels, the wall display is 4.1 meters (7710 pixels) wide and 2.31 meters (4350 pixels) high. The wall display is driven by a dual CPU (32 cores, 3.30 GHz) computer with four Nvidia Quadro K5000 cards. The two pointing devices used were a 4[th] generation iPad and a Vive controller. The controller is driven by another dedicated computer running Windows 10. The pointing data were streamed from the pointing devices to the display-driving computer over a local area network, for which the latency is measured to be within 2 ms for 90% of the packets. The experiment software was implemented as a web application, which recorded clicks in real time.

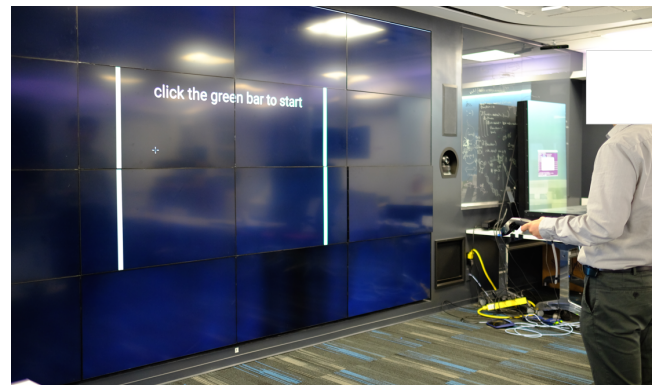

**Figure 2. The wall display used in the experiment, the stimuli, and the pointing posture.**

### Task

The task is divided into many *set*s of trials, in which each set involved alternatingly clicking two vertical bars on each side of the screen in succession for seven times (the first click starts the timer), with the target bar colored as green, and the non-target bar colored as white. The bar width and distance were maintained the same within a set, but varied across sets. The participants were asked to hold the controller in their dominant hand with their upper arm roughly vertical and forearm extended, as depicted by Figure 2. For the relative pointing condition, the participants held the iPad in their non-dominant hands, and used their dominant hands to control the cursor. The participants stood at a fixed location where when their forearms were extended, the controller was about two meters away from the screen.

The participants were told to complete the task as fast as possible while maintaining above 90% accuracy. When a click occurred outside the target bar, the target would flash red to indicate an error, and that trial was then excluded from the reaction time analysis, though the participant still had to click on the same target bar before proceeding to the next trial.

After the task, the participants completed the ISO 9241-9 survey [4], which includes 13 questions on a 5-point Likert scale about the operation, fatigue, and general usability of the four pointing techniques.

**Design**

The experiment had a 4x3x3 design. The three factors were the pointing *technique*, the target *width*, and the movement *amplitude*. As previously discussed, four pointing techniques were examined: relative, absolute, hybrid, and dual-speed. The target width was one of 25, 50, and 100 pixels, and the movement amplitude was chosen from 1000, 3000, and 5000 pixels. The resultant index of difficult values ranged from 3.46 to 7.65.

The three factors were fully crossed and were all within-subject variables. Because frequently switching between the pointing techniques may cause confusion, trials of the same pointing technique were grouped together into blocks, and the target width and amplitude levels were varied within the blocks. The order of the target widths and amplitudes were counterbalanced across participants using a balanced Latin square, while the order of the pointing techniques were randomized.

At the beginning of the experiment, the participants were introduced to the four pointing techniques and completed a set of practice trials for each technique. Then, before each block, the participants were asked to perform some more practice trials with the pointing technique used in the ensuing block. These practice trials always had a movement amplitude of 3000 pixels, and the target width alternated across sets between 25 pixels and 50 pixels. The participants had to keep practicing until they reach 90% accuracy over the last two sets of trials. This practice regime ensured that the participants were sufficiently trained before performing the formal trials.

**RESULTS**

**Error Analysis**

When collapsed over all levels of width and amplitude, on average, participants reached above 90% accuracy across all techniques, suggesting that the participants were indeed sufficiently trained over all pointing techniques. However, the mean accuracy for absolute pointing (90.8%) was still substantially lower than the other three techniques (relative: 95.8%, hybrid: 96.6%, dual-speed: 96.2%). Figure 3 and Figure 4 suggest a likely explanation for absolute pointing's lower accuracy. Figure 3 shows that the accuracy of absolute pointing was much more affected by the increase in ID than the other three pointing techniques. However, the drop in accuracy is not linear: It first dips at an ID of 5.36, followed by a steep rise, and then lowers again at IDs of 6.92 and 7.65. These three low accuracy points were all using the smallest target width, 25 pixels, suggesting that the accuracy of absolute pointing was mainly affected by target width. This observation is supported by Figure 4, and confirmed by a generalized mixed effects modeling analysis, which shows that pointing technique and width both had substantial influence on accuracy ($\lambda_c$ is 3.39 for pointing techniques and 4.74 for target width[2]), whereas all other factors and interactions had much less influence ($\lambda_c < 0.0006$).

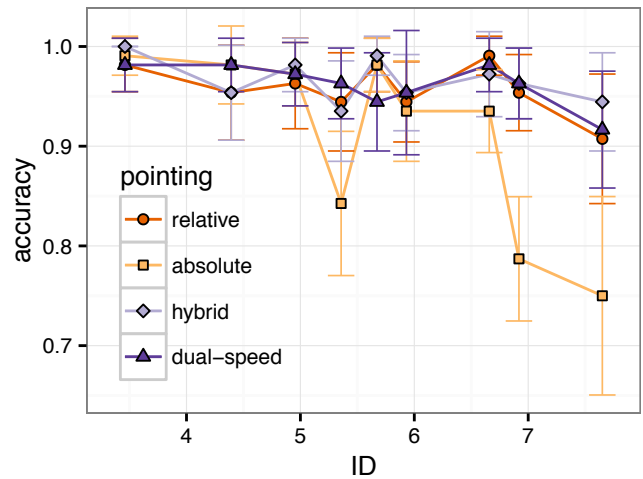

Figure 3. The participants' mean task accuracy as a function of index of difficulty (ID) across the four techniques. Error bars indicate 95% confidence intervals (CIs).

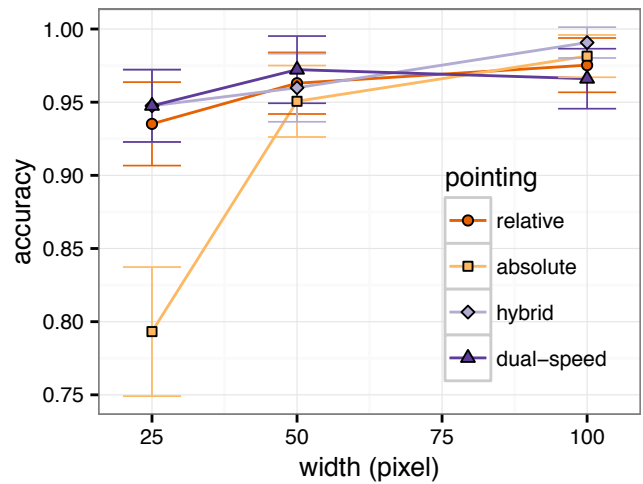

Figure 4. The participants' mean task accuracy as a function of target width. Error bars indicate 95% CIs.

**Reaction Time Analysis**

The time between two successive mouse down events that both landed on target bars was taken as the reaction time (RT), which ensured that only full-amplitude movements were analyzed. Each set involved six trials, and any error invalids a trial. To reduce the effect of outliers, which mainly resulted from participants' distraction and the occasionally false detection of fingers on the controller's trackpad, the median of the RTs of each set was taken as the final RT measure for each condition.

---

[2] $\lambda_c$ is the AIC corrected log likelihood ratio between unrestricted models and restricted models. It can be roughly interpreted as bits of evidence. Higher $\lambda_c$ suggests more influence of the independent variable on the dependent variable.

An ANOVA analysis shows a significant main effect of pointing technique, $F(1.98, 33.61)^3 = 23.16$, $p < .0001$. A Tukey's post-hoc comparison shows that hybrid was significantly slower than all other techniques (all $p$'s < .001), while dual-speed was significantly faster than absolute ($p = .036$) and relative ($p = .034$). These effects can also be seen in Figure 5, which shows the mean RTs of the four techniques collapsed over all levels of width and amplitude.

As expected, the ANOVA analysis also shows significant main effects of width ($F(2, 34) = 103.66$, $p < .0001$) and amplitude ($F(1.5, 25.54) = 124.86$, $p < .0001$). Participants were faster with wider bars and shorter movement distances. The interaction between pointing technique and width ($F(6, 102) = 9.43$, $p < .0001$) and the interaction between technique and amplitude ($F(2.56, 43.56) = 3.86$, $p = .02$) were also significant. All other interactions were non-significant.

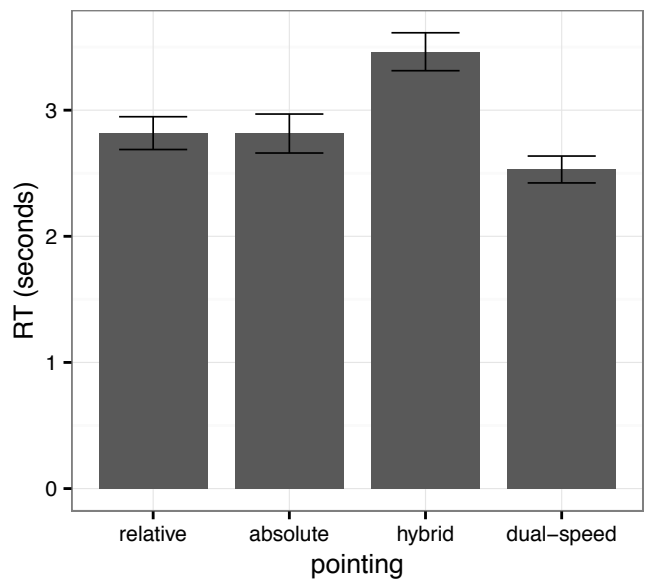

**Figure 5. The mean RTs of the four pointing conditions. Error bars indicate 95% CIs.**

*Analysis based on Fitts' ID*

For pointing tasks, using Fitts' index of difficulty (ID) rather than width and amplitude as the independent variable (IV) can vastly simplify the analysis. When using width and amplitude as IVs, they have to be coded into many dummy variables to model their nonlinear effects on RT. Fitts' ID captures the nonlinear effect directly, and therefore can be used in ANOVA as a single interval IV.

---

[3] Non-integer degrees of freedom indicate that Greenhouse-Geisser correction was applied due to violation of the assumption of sphericity.

The resultant, simplified model would facilitate the comparison of pointing techniques and the interpretation of the interactions between techniques and stimulus conditions.

Figure 6 shows RT as a function of ID for each pointing technique. As expected, for all pointing techniques, the main trend is that RT increases with ID. For all but the hybrid pointing technique, the participants' mean RTs fall close to the regression lines (dashed lines), suggesting that their data can be well explained by Fitts' law. This can also be seen in Table 1. The fitted Fitts' parameters for each pointing technique, and the goodness-of-fit measured as root-mean-squared error (RMSE) and $R^2$., in which only the hybrid pointing technique has an $R^2$ below 0.8, while others have $R^2$s above 0.92.

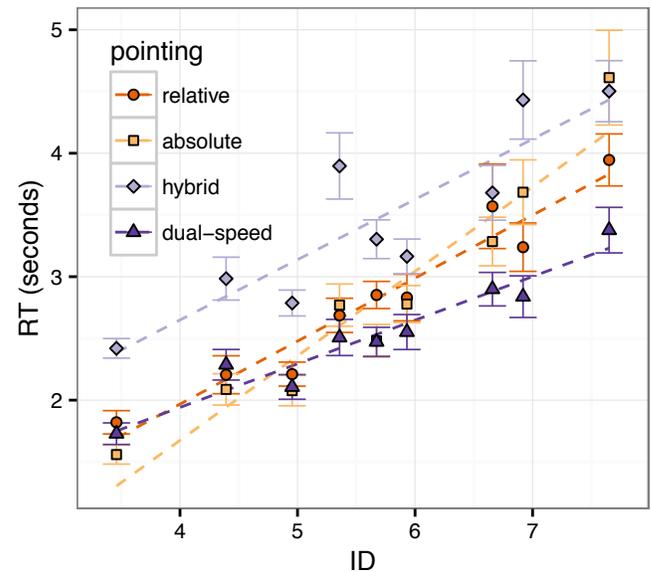

**Figure 6. Participants' mean RTs as a function of ID across the four pointing techniques. The dashed lines are the fitted regression lines. Error bars indicate 95% CIs.**

|  | *a* | *b* | *RMSE* | $R^2$ |
|---|---|---|---|---|
| relative | -0.074 | 0.510 | 0.155 | 0.943 |
| absolute | -1.064 | 0.685 | 0.253 | 0.918 |
| hybrid | 0.695 | 0.488 | 0.314 | 0.786 |
| dual-speed | 0.528 | 0.353 | 0.119 | 0.930 |

**Table 1. The fitted Fitts' parameters for each pointing technique, and the goodness-of-fit measured as root-mean-squared error (RMSE) and $R^2$.**

Because relative, absolute, and dual-speed pointing can be approximated by Fitts' law, their differences can be described by their Fitts' parameters: the intercept *a* and the slope *b*. A linear mixed effects model using the data from all but the hybrid conditions shows that though the intercept of dual-speed pointing is significantly higher than absolute ($p < .0001$) and relative pointing ($p = .042$), its slope is significantly smaller than theirs ($p < .0001$ for absolute, and $p = .001$ for relative). As a result, when the pointing task was easy (ID < 5), dual-speed pointing was slower than the other two techniques, but when the task was hard (ID > 5), dual-speed pointing was faster.

**Qualitative results**

Participants completed the ISO 9241-9 survey, which includes 13 questions assessing various aspects of the pointing techniques. This section reports data for four questions whose ratings were substantially different across the techniques. The rating mean and standard deviations for these four questions can be found in Table 2.

When asked about the mental effort required for the pointing operation (Question 3), participants rated relative pointing to be the least mentally demanding technique and hybrid pointing to be the most demanding technique. A Tukey pairwise post-hoc comparison shows that relative pointing required significantly lower mental effort than absolute ($p = 0.01$) and hybrid pointing ($p < .001$). The result is perhaps partly because participants were more familiar with the commonly used relative pointing technique, and partly due to the mode switches required by hybrid pointing. However, the difference between dual-speed and relative pointing is nonsignificant ($p = 0.1$), suggesting that cost of mode switches of dual-speed pointing is relatively lower than that of hybrid pointing.

|  | relative | absolute | hybrid | dual-speed |
|---|---|---|---|---|
| Q3. Mental effort | 1.56 (0.7) | 2.61 (1.46) | 2.94 (1.21) | 2.39 (1.24) |
| Q5. Accurate pointing | 2.11 (1.08) | 3.56 (1.38) | 2.28 (0.96) | 2.06 (0.8) |
| Q7. Finger fatigue | 2.67 (1.14) | 1.39 (0.61) | 2.11 (1.13) | 1.67 (0.69) |
| Q13. General usability | 3.78 (0.88) | 2.78 (1.26) | 3.61 (1.2) | 4.17 (0.79) |

**Table 2. The means and standard deviations, shown in parentheses, of the participants' ratings on four questions. Q3: A rating of 1 indicates very low mental effort, and 5 indicates very high mental effort. Q5: A rating of 1 indicates that accurate pointing was very easy, and 5 indicates very difficult. Q7: A rating of 1 indicates no finger fatigue, and 5 indicates very high finger fatigue. Q13: A rating of 1 indicates very difficult to use, and 5 indicates very easy to use.**

Participants reported that accurate pointing (Question 5) was significantly harder with absolute pointing than with all other pointing techniques (all $p$'s < .001). This is consistent with the quantitative results that showed absolute pointing had much lower accuracy than all other techniques.

When asked about the fatigue that the pointing techniques imparted on the various parts of the body, including finger, wrist, arm, shoulder, and neck, participants only rated finger fatigue (Question 7) substantially differently across the techniques. In particular, participants reported that relative pointing, which involved moving a finger on an iPad repetitively, caused significantly higher fatigue than absolute and dual-speed pointing (both $p$'s < .001). Several participants also commented that the iPad's glass surface is stickier than a typical trackpad surface, making it harder to drag a finger across long distances.

The last question (Question 13) of the survey asked about the general ease of use of each pointing technique, and the participants rated absolute pointing to be the most difficult one to use. Absolute pointing was found to be significantly more difficult to use than all other techniques ($p < .001$ when compared to relative; $p < .001$ for dual-speed, and $p = .007$ for hybrid). This result is probably due to the poor accuracy of absolute pointing. All participants commented that it was extremely difficult to use absolute pointing to click on small targets, and all but one participants had to use both hands to hold the controller in order to improve accuracy. Many of them even held the controller against their waist and rotated their whole upper body to move the cursor to gain accuracy. This suggests that in real world applications, pure absolute pointing is likely to encumber the users even if the controller only requires one hand to hold.

**DISCUSSION**

Consistent with prior studies [6,12,20], this research showed that the accuracy of absolute pointing is mainly affected by the target width. In addition, it seems the influence of target width on accuracy is nonlinear: It seems when the width was smaller than a threshold value, the error rate increased steeply. The threshold value was likely between 25 and 50 pixels, or 0.37° and 0.73° when converted into angular width. It is reasonable to believe that the error rates are affected by the angular width of the target rather than its pixel width on the screen since the source of the error are hand tremors.

As expected, the two dual-mode pointing techniques significantly improved the pointing accuracy over absolute pointing. Hybrid pointing improved accuracy by disassociating click actions from cursor position control, thus completely eliminated the adverse effect of hand tremors. Two participants also commented on this aspect of hybrid pointing, and thought it was very useful. Dual-speed pointing improved accuracy by reducing the impact of hand movements roughly by a factor of three (since the CD gain was 0.3), and it seemed sufficient for the smallest target width used in this study. In real world applications, the CD gain can be further reduced to accommodate smaller targets and longer screen distances.

Another possible way to improve the accuracy of dual-speeding pointing is to use the trigger button rather than the trackpad for mode switching and clicking. This might improve accuracy because clicking the trigger causes movements along the direction of pointing, whereas clicking the trackpad causes movements perpendicular to the pointing direction. However, a similar study done on Wii Remote didn't find significant difference between using a trigger button and a button on the top [9].

This study might be the first one to show that dual-speed pointing can be significantly faster than well-tuned relative pointing while achieving the same level of accuracy. Using 1/b as the throughput measure (as suggested by [21]), dual-speed had the highest throughput of all four techniques. However, when the task was easy, dual-speed pointing was about as fast as other techniques. This was due to its higher intercept parameter. The intercept parameter is a mixture of the constant components of a pointing task, which typically consists of time for recognizing the stimulus, time for clicking the button, and possibly time for switching pointing modes in the case of dual-mode pointing. Considering that both absolute and dual-speed pointing use the same pointing device and the same button for clicking and yet absolute pointing had much smaller intercept (see Table 1), the higher intercept of dual-speed pointing is likely caused by the mode switching cost. The same is true for hybrid pointing, which also had a high intercept value.

Surprisingly, hybrid pointing was the slowest technique. The combination of absolute pointing and a trackpad-like relative pointing had been tested in [11,20], and in both studies, it was as fast as pure relative pointing. This discrepancy might be caused by the different ways in which the relative pointing part operated. In this study, the trackpad of the controller is rather small, and it can only be operated by the thumb. In the other two studies, the trackpad was a large surface and was either operated by the index finger or by the entire hand. It might be that participants were still not used to using their thumbs to operate a trackpad. Despite its lower efficiency, two participants still commented that they liked hybrid pointing the most since there was no tremor at all, and one of them said "probably given more practice, I would be better at it".

Like [11,20], relative pointing is again shown to be an efficient pointing method even for large displays, as long as the CD transfer function parameters are well tuned. One practical advantage of the implementation here is that the pointing program is implemented as a webpage that can turn any touch devices into a trackpad without arduous installation processes. Since many people today carry a touch-enabled smartphone, this could be a good method to quickly enable many people to interact with large displays. However, relative pointing still has its inherent drawbacks when used with large displays such as the difficulty to find the cursor and the unclear mapping when there are displays at all directions. Thus, devices that enable some form of absolute pointing are still likely to be easier to use.

## CONCLUSIONS

This research proposed an inexpensive, long-range, and accurate pointing system based on HTC Vive. Two dual-mode pointing techniques that combine absolute and relative pointing were implemented using this new pointing device. These techniques were compared against pure absolute and pure relative pointing. The dual-speed pointing technique was found to be faster than all other techniques while also being one of the most accurate method. Given its low cost and robustness, this pointing system can be easily deployed into classrooms, meeting rooms, and homes for rich interactions and utilizations of large displays. The dual-speed pointing technique enables people to overcome the inherent problems in absolute pointing and efficiently interact with display-rich environments.